
\magnification=1200
\headline{\ifnum\pageno=1 \nopagenumbers
\else \hss\number \pageno \fi\hss}
\footline={\hfil}
\parindent=12pt
\baselineskip=20pt
\hsize=15 truecm
\vsize=23 truecm
\hoffset=0.7 truecm
\voffset=1 truecm
\overfullrule=0pt
\centerline{\bf STRANGENESS ENHANCEMENT}\par
\centerline{\bf IN HEAVY ION COLLISIONS} \par
\vskip 5 truemm
\centerline{\bf A. Capella} \par
\vskip 2 truemm
\centerline{Laboratoire de Physique Th\'eorique et Hautes
Energies\footnote{*}{Laboratoire
associ\'e au Centre National de la Recherche Scientifique - URA 63}} \par
\centerline{Universit\'e de Paris XI, b\^atiment 211, 91405 Orsay Cedex,
France} \par

\vskip 3 truecm
\noindent \underbar{{\bf Abstract}} \par
The enhancement of strange particle production observed in nucleus-nucleus
collisions at CERN is
explained by the combined effect of the increase in the relative number of
strings containing strange
constitutents at the ends, and the final state interaction of co-moving
secondaries~: $\pi + N \to K +
\Lambda/\Sigma$.
 \par

\vskip 3 truecm
\noindent LPTHE Orsay 94-113 \par
\noindent December 1994
\vfill \supereject
An increase of the ratio $R$ of strange over non-strange particles from
nucleon-nucleon to
central nucleus-nucleus collisions, as well as its increase with centrality, is
now firmly established
by several CERN and AGS collaborations. In the framework of independent string
models such as the Dual
Parton Model (DPM) and Quark Gluon String Model (QGSM), an increase of $R$ can
be partially
understood in the following way$^{[1][2]}$. When the average number of
collisions per participant
nucleon increases, the excess of produced particles is due to the fragmentation
of strings involving
sea constituents at their ends. Since strange quarks are present in the nucleon
sea, the ratio of
strange over non-strange secondaries will also increase. The size of this
increase depends on
the ratio $$S = {2(s + \overline{s}) \over u + d + \overline{u} +
\overline{d}}$$

\noindent in the nucleon sea. It turns out that the value $S = 0.5$ used in
conventional
parton distributions for DIS$^{[3]}$ allows to un\-ders\-tand the observed
enhancement of $K^-$, but
not those of $K^+$ and $K_s^0$.
Moreover, the enhancement of $\Lambda$ and $\overline{\Lambda}$ is negligeably
small.
Two approaches have been proposed in order to overcome the latter problem. In
DPM, diquark-antidiquark
pairs have been introduced in the nucleon sea - with the same ratio, relative
to
quark-antiquark pairs, as in the string breaking process$^{[2]}$. The second
approach, introduced by
several authors, is string fusion$^{[4, 5]}$. Both mechanisms have a common
caveat, namely, they
lead to the same increase in \underbar{absolute value} of strange baryons and
antibaryons  - which is
ruled out by ex\-pe\-ri\-ment. \par

In the following, we shall see that the introduction of diquarks in
the nucleon sea mentioned above allows, with the help of simple quark counting
arguments,
to account for the observed enhancement of $\overline{\Lambda}$ and
$\overline{\Xi}$. However, the
multiplicity of produced $\Lambda$'s is too small - by more than a factor of
two. The missing numbers
of $\Lambda$'s and kaons are approximately the same. Thus, some other
me\-cha\-nism is needed which
increases the number of $\Lambda$'s without increasing the number of $K^-$ and
$\overline{\Lambda}$.
A simple and well known way to do so, is by a final state interaction $\pi + N
\to K +
\Lambda/\Sigma$. Since at CERN energies the observed enhancement is very
important at mid-rapidities,
the final state interaction cannot be an intranuclear cascade. One needs a
genuine final state
interaction of co-moving secondaries. This interaction also plays a crucial
role in hadronic gas
models$^{[6]}$ and has been incorporated into several Monte Carlo
codes$^{[5][7][8]}$.\par

In this note I will show that the observed enhancement of strange mesons
and baryons/antibaryons can be understood by the combined effect of strange sea
quarks and
antiquarks and the final state interaction mentioned above. For simplicity, I
shall concentrate on the
strangeness enhancement from $NN$ to central $SS$ collisions. In order to
describe the
underlying physics in the most transparent way, I consider a simplified version
of the
DPM with only two strings for each $NN$ collision and make some simplifying
assumptions in the
fragmentation functions (see below). The aim is not to reproduce the data in
the best possible way,
but to present a plausible and, hopefully, convincing scenario of strangeness
enhancement
consistent with the experimental findings.   \par

As a further simplification, I take the configuration with an average number of
$\overline{n}_A = 24$
participant nucleons in each nucleus as representative of a central $SS$
collision. In this case the
average number of collisions is $\overline{n} = 53$\footnote{*}{These average
values are obtained in
the DPM based IRIS Monte Carlo, for events with $n_A$ and/or $n_B \geq 27$. I
thank
J. P. Pansart for giving me access to his code and A. Kouider for running it.}.
The DPM formula for
the average multiplicity of particle $i$ in $AA$ collisions is then$^{[9]}$
$$n_i = 2 \overline{n}_A
\ N_i^{(qq)_v-q_v} + 2 ( \overline{n} - \overline{n}_A) \
N_i^{q_s-\overline{q}_s} \eqno(1)$$

\noindent where $N_i^{(qq)_v-q_v}$ and $N_i^{q_s-\overline{q}_s}$ are the
corresponding average
multiplicities of valence $qq$-$q$ and sea $q$-$\overline{q}$ strings,
respectively. The total
number of strings is $2 \overline{n}$ (two for each $NN$ collision), out of
which
$2n_{\overline{A}}$ involve the valence quarks and diquarks of the
participating nucleons. The average multiplicity in $NN$ collisions is
$2N^{(qq)_v-q_v}$. Numerical calculations in
DPM show that the increase in the number of $\Lambda$ and $\overline{\Lambda}$
from $NN$ to $SS$ at
200 GeV/c per nucleon is roughly equal to half of the total average number of
participating nucleons - in our case 24. (The average mass of the
$q_s$-$\overline{q}_s$ strings is
too small to produce $\Lambda$-$\overline{\Lambda}$ pairs at CERN energies).
For kaons, this increase
turns out to be somewhat larger (approximately by a factor 28), due to the
kaons produced in
$q_s$-$\overline{q}_s$ strings. This enhancement factor increases with
decreasing va\-lues of the
mass of the produced particle. (This could explain the observed decrease of the
ratio $(n_{\rho} +
n_{\omega})/n_{charged}$ with increasing centrality$^{[10]}$). \par

The DPM results for the number of lambdas and kaons in $SS$ central collisions
at CERN can thus be
summarized as follows
 $$\eqalignno{
&n_{\Lambda(\overline{\Lambda})} = 24 n_{\Lambda (\overline{\Lambda})}^{NN}
&(2) \cr
&n_{K^{(i)}} = 24 n_{K^{(i)}}^{NN} + 4 n_K^{NN} &(3) \cr
}$$

\noindent where $n_i^{NN}$ are the corresponding multiplicities in $NN$
collisions at 200 GeV/c and
$n_K^{NN}$ is the average value over all kaon states. The second term of (3) is
the result of
expressing the contribution of the $q_s$-$\overline{q}_s$ strings in eq. (1) in
terms of the kaon
multiplicity in $NN$ at 200 GeV/c - hence the reduction in its coefficient from
$\overline{n}$-$\overline{n}_A$ = 29 to 4\footnote{*}{Computing the
contribution of the short
$q$-$\overline{q}$ strings at energies of a few hundreds of GeV/c is one of the
main sources of
numerical uncertainties in DPM. For this reason there is some uncertainty in
the kaon
multiplicities - which will also exist when a more sophisticated (Monte Carlo)
treatment is used.
This uncertainty is not present in $n_{\Lambda(\overline{\Lambda})}$.}.
Obviously these strings give
the same number of $K^-$, $K^+$ and $K_s^0$. Using the experimental
values\footnote{**}{The typical
errors in these numbers range from 10 $\%$ for $K_s^0$ and $\Lambda$ to 30 $\%$
for
$\overline{\Lambda}$. The corresponding errors in the $SS$ multiplicities are
easy to determine.
However, the errors due to the uncertainties in the parameters of the model are
much larger (see
below).} of $n_i^{NN}$ given in ref. [11]~:
$$n_{\Lambda(\overline{\Lambda})}^{NN} = 0.096 \ (0.013)\ , \quad
n_{K^-(K^+)}^{NN} = 0.17 \ (0.24)
\ , \quad n_{K_s^0}^{NN} = n_K^{NN} = 0.20 \eqno(4)$$

\noindent we get
$$n_{\Lambda} = 2.3 \ (9.4 \pm 1) \ , \quad n_{\overline{\Lambda}} = 0.31 \
(2.2 \pm 0.4) \eqno(5)$$

$$n_{K^-} = 4.9 \ (6.9 \pm 0.4) \ , \quad n_{K^+} = 6.6 \ (12.4 \pm 0.4) \ ,
\quad n_{K_s^0} = 5.6 \
(10.5 \pm 1.7) \ \ \ . \eqno(6)$$

\noindent These numbers are much smaller than the experimental ones$^{[12]}$
given in bra\-ckets. The
numerical results (2)-(6) are obtained considering only $u$ and $d$ quarks in
the nucleon sea. \par

 When also strange quarks are present, eq. (1) can be written as
$$n_i = 2 \overline{n}_A \ N_i^{(qq)_v-q_v} + 2 (n - \overline{n}_A) \left ( {S
\over 2 + S}
N^{q_s^S-\overline{q}_s^S} + {2 \over 2 + S} N_i^{q_s^{NS}-\overline{q}_s^{NS}}
\right ) \eqno(7)$$

\noindent where the contribution of strange $(q^S)$ and non-strange $(q^{NS})$
sea quarks are
explicitly given (sea quarks have, of course, to be linked to sea antiquarks in
all possible ways). We
see from eq. (7) that the number of kaons will increase as a result of the
fragmentation of
$q^S$-$\overline{q}^S$ strings - since the fragmentation
$s(\overline{s}) \to K^-(K^+)$ is larger than the corresponding one for
non-strange quarks. We
take this fragmentation function to be the same as $d \to \pi^-$ $(u \to
\pi^+)$, since in both cases
one has to pull out a $u\overline{u}$ ($d\overline{d}$) pair in the first
string
break-up. At CERN energies, the average mass of the $q$-$\overline{q}$ strings
is so small that
further break-ups are not possible. The same arguments leading from eq. (1) to
(3), lead now from eq.
(7) to   $$n_{K^{(i)}}
= 24 n_{K^{(i)}}^{NN} + 4 \left ( {S \over 2 + S} n_{\pi}^{NN} + {2 \over 2 +
S} n_{K}^{NN}
\right ) \eqno(8)$$

\noindent which reduces to (3) for $S = 0$. Here $n_{\pi}^{NN} = 3.04$ is the
pion multiplicity in nucleon-nucleon averaged over its three charge
states$^{[11]}$. The first
(second) term inside the bracket corresponds to the contribution of
$q^S$-$\overline{q}^S$
$(q^{NS}$-$\overline{q}^{NS})$ strings. The resulting values of the kaon
multiplicities are now
$$n_{K^-} = 7.1 \ , \quad n_{K^+} = 8.8 \ , \quad n_{K_s^0} = 7.9 \ \ \ .
\eqno(8')$$

\noindent The multiplicity of $K^-$ is now in agreement with experiment - while
those of $K^+$ and
$K_s^0$ are still too low by about three units each. This difference is quite
significant and cannot
be explained in the framework of independent string models. Comparing (6) with
(8'), we see that
the strange sea quarks have produced an increase of 40 $\%$ in the kaon average
multiplicity. For
$SU$ collisions the corresponding increase is almost a factor 2. This could
perhaps explain
the enhancement of the ratio $n_{\phi}/(n_{\rho} + n_{\omega})$ observed
experimentally$^{[10][13]}$
in central $SU$ collisions. \par

Let us turn to $\Lambda$ and $\overline{\Lambda}$ production. As already
discussed, $q$-$\overline{q}$
strings cannot produce $\Lambda$-$\overline{\Lambda}$ pairs at CERN energies,
and one has to
introduce diquark pairs in the nucleon sea to account for its production.
Following [2], I assume that
their relative fraction $\alpha$ is the same as in the string breaking process,
and I take the value
$\alpha \sim 0.1$ from the JETSET code$^{[14]}$. To first order in $\alpha$,
eq. (1) is then changed
into$^{[2]}$~: $$n_i = 2
\overline{n}_A \ N_i^{(qq)_v-q_v} + (\overline{n} - \overline{n}_A) \left [ (1
- 2 \alpha )
2N_i^{q_s-\overline{q}_s} + 2 \alpha \left ( N_i^{(qq)_s-q_s} +
N_i^{(\overline{q}\overline{q})_s-\overline{q}_s} \right ) \right ] \ \ \ .
\eqno(9)$$

\noindent The last term in (9) corresponds to the contribution of strings with
sea
diquarks\footnote{*}{It is easy to see that when using eq. (9) instead of eq.
(1) (or eq. 7) the
number of kaons is practically unchanged.}. It will produce a substantial
number of
$\Lambda$-$\overline{\Lambda}$ pairs when the sea diquarks are either $us$ or
$ds$. I shall take the
fragmentation functions $us$ ($ds$) $\to \Lambda$ to be the same as $uu$ ($ud$)
$\to p$ - since in
both cases one has to pull out a $d\overline{d}$ ($u\overline{u}$) pair in the
first string break-up
which produces the leading baryon. The number of $\Lambda$-$\overline{\Lambda}$
pairs coming from the
last term of eq. (9) is        $$\Delta n_{\Lambda(\overline{\Lambda})} =
\alpha (\overline{n} -
\overline{n}_A) \left [ {4S \over 4 + 4S + S^2} n_p^{pp} + {2 \over 4 + 4S +
S^2} \ {3 \over 2}
n_{\Lambda}^{pp} \right ] = 1.4 \ \ \ . \eqno(10)$$

\noindent with $n_p^{pp} = 1.34$ and $n_{\Lambda}^{pp} = 0.096^{[11]}$. The
first term of eq. (10),
which gives the most important contribution to $\Delta n$, is the result of the
fragmentation of the
strings with $us$, $su$, $ds$ or $sd$ diquarks at one end. The corresponding
probability is $4S$
divided by the total weight $4 + 4S + S^2$ - where 4 is the weight of the non
strange diquarks $uu$,
$dd$, $ud$ and $du$ and $S^2$ is the one of the $ss$ diquark. As explained
above the average
$\Lambda$ multiplicity is in this case equal to $n_p^{pp}$. The second term
corresponds to the
fragmentation of strings containing $ud$ and $du$ diquarks\footnote{*}{Only
diquark fragmentations
that occur without diquark breaking have been included in eqs. (10)-(12). The
factor ${3 \over 2}$ in
the second term of (10) and (11), is due to the fact that only two (out of the
three diquarks of the
proton) can fragment into $\Lambda$ without breaking. Almost identical results
are obtained when the
``pop-corn'' diquark breaking mechanism is taken into account. For instance,
the contribution to the
second term of eq. (10) of $uu$ ($dd$) $\to \Lambda$ diquark breaking
fragmentation, is approximately
compensated by a corresponding decrease of the factor ${3 \over 2}$.}. \par

Adding (10) to (5), the $\overline{\Lambda}$ multiplicity is
$n_{\overline{\Lambda}} = 1.7$, which is
close to the experimental value - while $n_{\Lambda} = 3.7$ is still too low by
more than a factor 2
(about six units). \par

Likewise, the extra $\Xi^-$-$\overline{\Xi}^+$ pair production is given by
$$\Delta n_{\Xi(\overline{\Xi})} = \alpha (\overline{n} - \overline{n}_A)
\left ( {S^2 \over
4 + 4S + S^2} n_p^{pp} + {2S \over 4 + 4S + S^2} \ {3 \over 2} n^{pp}_{\Lambda}
\right ) = 0.22
\eqno(11)$$

\noindent The value of $n_{\overline{\Xi}}^{pp}$ has not been directly
measured. From the
ratio $n_{\overline{\Xi}}^{pp}/n_{\overline{\Lambda}}^{pp} = 0.06 \pm
0.02^{[15]}$, we deduce
$n_{\overline{\Xi}}^{pp} \simeq 0.001$. We then have    $$n_{\overline{\Xi}} =
24n_{\overline{\Xi}}^{pp} + \Delta n_{\overline{\Xi}} = 0.24 \ \ \ , \qquad
n_{\overline{\Xi}}/n_{\overline{\Lambda}} = 0.14 \ \ \ . \eqno(12)$$

\noindent Somewhat larger values of
$n_{\overline{\Xi}}^{pp}/n_{\overline{\Lambda}}^{pp}$ have
been measured in $e^+e^-$ and $\overline{p}p$ at high energies. A recent
measurement from the E735
coll.$^{[16]}$ suggests a value as large as 0.25. This places the ratio (12) in
the range 0.14 $\div$
0.18. Notice that this ratio is rather insensitive to variations of $S$ and/or
$\alpha$. It is also
practically independent of the process~: one gets 0.15 $\div$ 0.19 in $SAu$.
This value has to be
compared with the experimental numbers~: 0.20 $\pm$ 0.03 in $SW$ from the WA85
coll.$^{[17]}$ and
0.127 $\pm$ 0.022 in $SPb$ from NA36 coll.$^{[18]}$ (the latter in a limited
range of $y$ and
$p_{\bot}$). A similar value of this ratio was found in [2]. \par

The same argument allows to determine the extra number of
$\Omega$-$\overline{\Omega}$ pairs, which is
given by
$$\Delta n_{\Omega(\overline{\Omega})} = \alpha (\overline{n} - \overline{n}_A)
{S^2 \over 4
+ 4S + S^2} {3 \over 2} n_{\Lambda}^{pp} = 0.02 \ \ \ , \eqno(13)$$

\noindent leading to a ratio $n_{\overline{\Omega}}/n_{\overline{\Xi}} \sim
0.1$. \par

It is important to note that $n_p^{pp}$ and $n_{\Lambda}^{pp}$ in Eqs.
(10)-(13)
have to be taken at the reduced energy $\sqrt{s/3} \sim$ 11 GeV. This is due to
the presence of three
diquarks, which share most of the momentum of the nucleon. This is of little
consequence
for the average $SS$ multiplicities - since the values of $n_p^{pp}$ and
$n_{\Lambda}^{pp}$
measured at 69 GeV/c$^{[19]}$ are consistent with those at 200 GeV/c$^{[11]}$.
However, this fact is
crucial to understand the rapidity distributions ~: the
$\Lambda$-$\overline{\Lambda}$ pairs
will be concentrated in the region $|y^*| < Y_{MAX}^* \sim 2$, as observed
experimentally$^{[12]}$.
\par

As discussed above, the number of missing $\Lambda$'s is about six units and
those of
missing $K^+$ and $K_s^0$ are about three units each. The most obvious
mechanism that produces extra
$\Lambda$, $K^+$ and $K_s^0$, without changing the number of the other
particles, is the
final state interaction  $$\pi + N \to K + \Lambda/\Sigma \eqno(14)$$

\noindent where $\pi$ and $N$ are co-moving secondaries. It is clear that the
process (14) is very
favorable to produce extra $\Lambda$'s. Indeed its gain is proportional to the
density of
$N$ (much  larger than the one of $\Lambda$'s), times the density of pions -
which is
even larger. By strangeness conservation, kaons will be produced exactly in the
same amount
(equally shared between $K_s^0$ and $K^+$) - while $K^-$ and
$\overline{\Lambda}$ are not
produced. Notice that the losses of $K$'s and $\Lambda$'s resulting from the
crossed processes
$$\pi + \Lambda \to K + N \ \ (a) \quad , \qquad
K + N \to \pi + \Lambda \ \ (b) \ \ \ , \eqno(15)$$

\noindent with cross-sections comparable to the one of the direct process (14),
produce
small effects in the final balance of produced particles. For instance, the
loss of
$\Lambda$'s in (15a) relative to its gain in (14) is suppressed by the ratio of
$\Lambda$
over $N$ densities. Moreover the loss of $\Lambda$'s in (15a) and its
gain in (15b) tend to compensate one another. The same is true for the balance
of kaons. Therefore,
I shall concentrate on the gain of $K$'s and $\Lambda$'s resulting from
(14)\footnote{*}{Probably the real situation is more complicated with some
extra gain of kaons due
to $\pi \pi \to K \overline{K}$ and some extra loss of kaons (and gain of
strange baryons and
antibaryons) due to the strangeness exchange reactions ($KN \to \pi
\overline{Y}$ and
$\overline{K}N \to \pi Y$) which have a large cross-section at
threshold$^{[6][22]}$}. This gain is
proportional to the product of $\pi$'s and $N$'s densities times the
cross-section of process (14).
This cross-section, suitably averaged over the momentum distributions of the
colliding particles near
the production threshold, will be denoted by $<\sigma >$. Following Ref. [20],
we have \par
$${dn_{\Lambda} \over dy} = \int d^2s {dn_{\pi_-} \over dy d^2s} \ {dn_p \over
dy \ d^2s} 3<\sigma
> \ell n \left [ \tau + \tau_0)/\tau_0 \right ] \eqno(16)$$

\noindent where $\tau_0$ is the formation proper time, $\tau$ the time during
which the final state
interaction takes place and $d^2s$ a differential transverse area. (For a
collision at zero
impact parameter, $|\vec{s}|$ measures the distance between this differential
area and the axis
determined by the centers of two colliding nuclei). The factor three comes from
the product of three
pion times two nucleon species divided by a factor 2 which is due to the fact
that the
cross-sections for $\pi^+p$ and $\pi^-n$ are negligeably small. \par

For $SS$ collisions at $b \sim 0$, one gets by using Gaussian nuclear
profiles\footnote{*}{At fixed
$b$, one has~: $dn/dy \ d^2s \propto \overline{n}_A(b, s) \propto T_A(b - s)
\sigma_{NA}(s)$, where
$\sigma_{NA}(s) = 1 - (1 - \sigma T_A(s))^A$ and $T_A(b) = \exp (- b^2/r^2)/\pi
r^2$ with $r^2 = 2
R^2/3$.}, $${dn_{\Lambda} \over dy} = {3 <\sigma > \over \pi R^2} \ {dn_{\pi^-}
\over dy} \ {dn_p
\over dy} \ell n \left [ \tau + \tau_0)/\tau_0 \right ] \eqno(17)$$

\noindent where $R$ is the r.m.s. radius of Sulfur. I take $\tau_0 \sim 1$ fm
and $\tau + \tau_0
\sim \tau_f \sim$ 4 fm, based on interferometry measurements which indicate a
very short time of
particle emission and a freeze out time of 4 fm$^{[21]}$. For the value of
$<\sigma >$, I take
$<\sigma >$ = 1.5 mb. (The sum of $\pi N \to K \Lambda$ plus $\pi N \to K
\Sigma$
cross-sections at their maxima is 1.5 mb$^{[22]}$. Beyond the maximum these
cross-sections decrease
sharply but quasi two-body processes convert this sharp decrease into a mild
increase). Using the
experimental data of the NA35 coll.$^{[23]}$ for the $dn_p/dy$ and
$dn_{h^-}/dy$ rapidity densities
(the latter multiplied by $n_{\pi^-}^{NN}/n_{h^-}^{NN}$), one gets from eq.
(17) $$\Delta n_{\Lambda}
= 2 \Delta n_{K^+} = 2 \Delta n_{K_s^0} = 6.2 \ \ \ . \eqno(18)$$

\noindent A practically identical result is obtained using the values of the
$\pi^-$ and $p$
densities computed in DPM. \par

To summarize, adding the values obtained above, the average multiplicities of
strange
particles in $SS$ collisions at 200 GeV/c per nucleon are~:
$$n_{K^-} = 7.1 \quad , \qquad n_{K^+} = 11.9 \quad , \qquad n_{K_s^0} = 11.0
\ \ \ , $$
$$n_{\Lambda} = 9.9 \quad , \qquad n_{\overline{\Lambda}} = 1.7 \quad , \qquad
n_{\overline{\Xi}} =
0.24 \div 0.30 \ \ \ . $$

\noindent The ratio $n_{\overline{\Xi}}/n_{\overline{\Lambda}} = 0.14 \div
0.18$ is practically the
same in $S$-$S$ and $S$-$Au$ collisions. \par

In conclusion, the scenario of strangeness enhancement presented above is based
on two mechanisms~:
the presence of strange quarks and diquarks in the nucleon sea, controlled by
the parameters $S \sim
0.5$ and $\alpha \sim 0.1$, and the final state interaction of co-moving
secondaries $(\pi N \to K
\Lambda/\Sigma )$, controlled by the parameter $< \sigma > \ell n (\tau +
\tau_0)/\tau_0$. There
are uncertainties in the values of these parameters. The value $S \sim 0.5$ is
used in DIS$^{[3]}$
but there is also indirect evidence for such a large fraction of strange quarks
from $\pi N$
scattering$^{[24]}$. The value of the strangeness suppression factor in the
string breaking
process is believed to be somewhat smaller - a value 0.4 is suggested by the
$\Lambda$/$p$ ratio
at mid-rapidities measured at Fermilab$^{[16]}$. Using $S = 0.4$, instead of $S
= 0.5$, reduces the
final numbers of $\Lambda$ and $\overline{\Lambda}$ computed above by only 0.15
units - and those of
kaons by 0.4. There are also uncertainties in the precise formulation of the
final state interaction
as well as in the value of the parameter that controls it. Moreover, the above
scenario is consistent
only if all other rescattering processes$^{[6]}$ contributing to the gain and
loss of
strangeness in a hadronic gas approach give small enough
effects. \par

The introduction of diquarks in the nucleon sea has some common features with
string
fusion$^{[4][5]}$. It is, on the contrary, in sharp contrast with a phase
transition scenario at the partonic level (QGP), considered by many
authors$^{[25]}$ - where the
production rate of strange quarks increases in the deconfined phase$^{[22]}$.
Actually, in the
present approach the strange quarks and diquarks are an \underbar{intrinsic}
property of the nucleon
wave function and therefore the parameters $S$ and $\alpha$ are expected to be
universal, i.e.
independent of (the centrality of) the process and of its energy. A detailed
study of strange
particle production in various processes, including predictions for $PbPb$, is
in progress. It will
be very interesting to compare them with the forthcoming results on strangeness
enhancement from lead
beam experiments at CERN. \par

\vfill \supereject  \noindent{\bf
\underbar{Acknowledgements}} \par \smallskip  It is a pleasure to thank
A.~Kaidalov for interesting
discussions and suggestions on the role of final particle interaction. I also
thank A.~Bialas,
K.~Fialkowski, A.~Kouider, A.~Krzywicki, C.~Merino, J.~Ranft, C.~Pajares and
J.~Tran Thanh Van for
interesting discussions. An early discussion with J.~Casado on the introduction
of sea diquarks
in DPM is also acknowledged.

\vfill \supereject
\centerline{\bf \underbar{References}} \par \bigskip

\item{[1]} A. Capella, U. Sukhatme, C. I. Tan and J. Tran Thanh Van, Phys. Rep.
\underbar{236}
(1994) 225.
\item{[2]} J. Ranft, A. Capella, J. Tran Thanh Van, Phys. Lett. \underbar{B320}
(1994) 346~;
\item{} H. J. M\"ohring, J. Ranft, A. Capella, J. Tran Thanh Van, Phys. Rev.
\underbar{D47} (1993)
4146 (the calculations in these papers are based on the DPMJET and DTNUC
codes).
\item{[3]} A.D. Martin, R. G. Roberts and W. J. Stirling, preprint RAL-94-005
(DTP/94/3).
\item{[4]} N. Armesto, M.A. Braun, E.G. Ferreiro and C. Pajares, University of
Santiago de
Compostela, preprint US-FT/16-94. References to earlier papers on string fusion
can be found in G.
Gustafson, Nucl. Phys. A \underbar{566} (1994) 233c.
\item{[5]} RQMD~: H. Sorge, R. Matiello, A. von Kectz, H. St\"ocker and W.
Greiner, Z. Phys.
\underbar{C47} (1990) 629~; H. Sorge, M. Berenguer, H. St\"ocker and W.
Greiner, Phys. Letters
\underbar{B289} (1992) 6.
\item{[6]} P. Koch and J. Rafelski, Nucl. Phys. \underbar{A444} (1985)
678.
 \item{[7]} QGSM : A. Kaidalov, Phys. Lett. \underbar{B117} (1982) 459~; A.
Kaidalov and K. A.
Ter-Martirosyan, Phys. Lett. \underbar{B117} (1982) 247. For the corresponding
Monte Carlo code see
N. S. Amelin et al., Phys. Rev. \underbar{C47} (1993) 2299. \item{[8]} VENUS :
K. Werner, Nucl. Phys.
\underbar{A566} (1994) 477c~; Phys. Rep. \underbar{232} (1993) 87~; preprint
HD-TVP-94-6.
\item{[9]} A. Capella, C. Pajares, V.A. Ramallo, Nucl. Phys.
\underbar{B241} (1984) 75~; A. Capella, J. Kwiecinski, J. Tran Thanh Van, Phys.
Lett.
\underbar{B108} (1982) 347.
\item{[10]} HELIOS/3 coll. : A. Manzoni, Nucl. Phys. \underbar{A566} (1994)
95c.
\item{[11]} M. Gazdzicki and
Ole Hansen, Nucl. Phys. \underbar{A528} (1991) 754. \item{[12]} NA35 coll. : T.
Alber et al., Z.
Phys. \underbar{C64} (1994) 195. \item{[13]} NA38 coll. : C. Baglin et al.,
Phys. Lett. \underbar{B259}
(1991) 62~;   \item{} R. Ferreira, Nucl. Phys. \underbar{A544} (1992) 623c.
\item{[14]} B. Andersson, G. Gustafson and T. Sj\"ostrand, Physica Scripta
\underbar{33} (1985) 574~;
\item{} T. Sj\"ostrand, CERN-TH 6488/92.
\item{[15]} AFS coll. : T. {\AA}kesson et al., Nucl. Phys. \underbar{B246}
(1984) 1.
\item{[16]} E765 coll.~: T. Alexoupoulos et al., Phys. Rev. \underbar{D46}
(1992) 2773.
\item{[17]} WA85 coll. : E. Quercig, Nucl. Phys. \underbar{A566} (1994) 321c~;
\item{} D. Evans, ibid p. 225c.
\item{[18]} NA36 coll. : J. M. Nelson, Nucl. Phys. \underbar{A566} (1994) 217c.
\item{[19]} V. V. Amomsov et al., Nuovo Cimento \underbar{40A} (1977) 237.
\item{} L. H. Blumenfeld et al.,
Phys. Lett. \underbar{45B} (1973) 528. \item{[20]} P. Koch, U. Heinz and J.
Pitsut, Phys. Lett.
\underbar{B243} (1990) 149.
\item{[21]} NA35 coll : G. Roland, Nucl. Phys. \underbar{A566} (1994)
527c~; D. Ferenc, Proceedings 29th Rencontres de Moriond (1994) ed. J. Tran
Thanh Van~;
\item{} NA44
coll. : T.J. Humanic, Nucl. Phys. \underbar{A566} (1994) 115c~; S. Panday,
Proceedings
Rencontres de Moriond, ibid. \item{[22]} B. Koch, B. Muller and J. Rafelski,
Phys. Rep.
\underbar{142} (1986) 167. A compilation of $\pi N \to K \Lambda/\Sigma$ and
$KN \to
\pi \Lambda$ cross-sections is given in Appendix B.   \item{[23]} NA35 coll. :
J. Bartke et al., Z.
Phys.  \underbar{C48} (1990) 191~; \item{} D. R\"ohrich, in Nucl. Phys.
\underbar{A566} (1994) 35c.
\item{[24]} B. Ioffe and M. Karliner, Phys. Lett. \underbar{B247} (1990) 387
and references therein.
\item{[25]} J. Lettessier, A. Tounsi, U. Heinz, J. Sollfrank and J. Rafelski,
Phys. Rev.
Lett. \underbar{70} (1993) 3530 and preprint PAR/LPTHE/92-27R~; N. Heinz, Nucl.
Phys. \underbar{A566}
(1994) 205c~; K. Redlich, J. Cleymans, H. Satz and E. Suhonen, Nucl. Phys.
\underbar{A566}
(1994) 391c and references therein~; M. Ga\'zdzicki and St. Mr\'owczy\'nski, Z.
Phys. \underbar{C49}
(1991)~; H. W. Barz, B. L. Friman, J. Knoll, H. Schulz, Nucl. Phys.
\underbar{A484} (1988)
661~; K. Geiger, Phys. Rev. \underbar{D48} (1993) 4129. See also ref. [7].

 \bye